\documentclass{sig-alternate-2013}

\usepackage[
  pass,
]{geometry}

\newfont{\mycrnotice}{ptmr8t at 7pt}
\newfont{\myconfname}{ptmri8t at 7pt}
\let\crnotice\mycrnotice%
\let\confname\myconfname%
\permission{Permission to make digital or hard copies of all or part of this work for personal or classroom use is granted without fee provided that copies are not made or distributed for profit or commercial advantage and that copies bear this notice and the full citation on the first page. Copyrights for components of this work owned by others than ACM must be honored. Abstracting with credit is permitted. To copy otherwise, or republish, to post on servers or to redistribute to lists, requires prior specific permission and/or a fee. Request permissions from Permissions@acm.org.}
\conferenceinfo{DocEng'15,}{September 08 - 11, 2015, Lausanne, Switzerland.}
\copyrightetc{\copyright~2015 ACM. ISBN\the\acmcopyr}
\crdata{978-1-4503-3307-8/15/09\ ...\$15.00.\\
DOI: http://dx.doi.org/10.1145/2682571.2797090}

\usepackage{hyperref}
\usepackage{subfigure}
\usepackage{multicol}
\usepackage[font=normalsize,skip=0pt]{caption} 
\usepackage{array}
\usepackage{multirow}
\usepackage{float}
\restylefloat{table}
\begin{document}
\title{MSoS: A Multi-Screen-Oriented Web Page Segmentation Approach}
\numberofauthors{1}
\author{
\alignauthor
Mira Sarkis, Cyril Concolato, Jean-Claude Dufourd\\
\affaddr{Telecom ParisTech; Institut Mines-Telecom; CNRS LTCI}\\
\email{\{sarkis, concolato, dufourd\}@telecom-paristech.fr}
}
\maketitle
\begin{abstract}
In this paper we describe a multiscreen-oriented approach for segmenting web pages.
The segmentation is an automatic and hybrid visual and
 structural method. It aims at creating coherent blocks which have different functions determined by the multiscreen environment. It is also characterized by a dynamic adaptation to the page content. 
Experiments are conducted on a set of existing applications that contain multimedia elements, in particular YouTube and video player pages.  Results are compared with one segmentation method from the literature and with a ground truth manually created. With a 81\%\ precision, the MSoS is a promising method that is capable of producing good segmentation results.
\end{abstract}


\section{Introduction}
Understanding and analyzing web content at the Internet scale requires automatic processing techniques. These techniques try to simulate the human understandability in terms of visualization, semantic meaning and interaction.
Among the existing techniques for web content analysis, web page segmentation techniques are widely used. They consist in decomposing a page into blocks that englobe coherent and related content. Segmentation is used in the adaptation of content to mobile, printing devices or in applications performing information extraction, among others.

In a multi-screen environment, where multiple devices are used to display and to interact with related content, users want to have their applications distributed among their devices. For instance, using the touch-screen of a smart-phone to control the smart-TV functions, while the large screen of the smartTV displays the multimedia content. In order to efficiently exploit the features of each device, the distribution is achieved by splitting the application into multiple pieces and by associating each piece to a proper device. The challenges here are: (1) to identify coherent blocks of content that can be separated from the rest without breaking the web page structure, (2) to know the device features wherein the content is efficient for the end-user and (3) to automatically map the content blocks to the 'best-match' device.

With this motivation, this paper has one principal objective: \emph{To propose a segmentation method that is automatic and guided by the multi-screen environment, based on: (1) visual analysis, (2) DOM analysis, and 
(3) analysis of content functions in order to achieve the application distribution in a multiscreen environment.}
A content function refers to the type of interaction between a user and a block of content, e.g., 'display' for multimedia content and 'interaction' for interactive content. In contrast to existing works, our intention is to guide the segmentation based on the features of target devices to facilitate the mapping of blocks to devices. 
We call our approach multiscreen-oriented segmentation and we refer to it as MSoS.
The validation of MSoS is performed in the virtual splitting system \cite{theVirtualSplitter} that re-factors web pages to create multi-screen applications and hereinafter referred to as VSplitter. 
We have tested MSoS on a set of pages featuring video elements and interactive content, e.g., YouTube pages and video-player applications. 
Through experimentation, we show how the MSoS adapts to the page content to produce better results when compared to Block-o-Matic\cite{bom}, a method in the literature and to a ground truth.

This paper is organized as follows. Section \ref{sec:SOA} introduces our MSoS approach within the state-of-the-art. In Section \ref{sec:modifiedBoM} MSoS is described. The implementation and the evaluation of MSoS in the VSplitter are described in Section \ref{sec:experimentation}. Finally conclusions are drawn in Section \ref{sec:conclusion}.

\section{State of the art}
\label{sec:SOA}

\textbf{Segmentation techniques:}
Hybrid segmentation techniques can get better results compared to techniques that are based only on one type of analysis, i.e., DOM, visual or content analysis. For instance the hybrid VIPS \cite{vips}, based on the joint DOM and visual analysis, utilizes both structural information in the DOM tree and visual cues to semantically segment a page. The hybrid Block-o-Matic (BoM) platform \cite{bom}, based also on the joint DOM and visual analysis, additionally abstracts the segmentation from the DOM tree and works at higher levels. This abstraction facilitates the understanding and the processing of the page structure. BoM starts by filtering the DOM structure based on the W3C content classification\footnote{http://www.w3.org/TR/html5/dom.html\#content-categories} and on the geometrical features to form a logical tree. Afterwards, the logical tree is processed based on Gestalt laws, i.e., proximity, similarity, closure and simplicity, and on the degree of granularity for merging nodes. At the end of the segmentation, the final blocks are represented in the logical tree by leaf nodes that mainly contain information about the node geometry and the corresponding DOM elements. This link between logical nodes and DOM elements makes the segmentation results easily exploitable by other applications. 

Though the processing of BoM is totally automatic, its configuration with a granularity parameter (pG) is manual and has to be tailored for each page. The pG value determines the threshold under which a logical node automatically produces a block and above which the node's children are processed individually. This value dictates the segmentation results. Configuring BoM with an inadequate pG value leads to a page not correctly segmented, and applying BoM with the same pG value on a heterogeneous page does not always create coherent blocks similarly on the whole page.

\textbf{Identifying block functions:}
BoM does not separate the blocks based on their functions, but it labels blocks with labels that are not relevant for our multi-screen environment, e.g., header, content, image, logo, etc.\cite{bom}. 
A function-based object model (FOM) for website adaptation is introduced by Chen et al.\cite{FOM}. The segmentation model defines a block as a set of information that have a specific function, i.e., information, navigation, interaction, decoration or others. In FOM, even if a function reflects the intention of the author for using this object, it does not reflect the type of interaction with the end-user. 

\textbf{Positioning our approach:}
In our work, we reuse the hybrid approach and the abstraction model proposed by BoM but we adapt the segmentation to make it completely automatic and multiscreen-oriented. We propose in particular to update the pG value based on the content. Additionally, our approach reuses the idea of identifying the block functions from the page content as in FOM, but we define functions from the end-user perspective and not the author. 

\begin{figure}
\centering
\includegraphics[height=0.6in,width=0.45\textwidth]{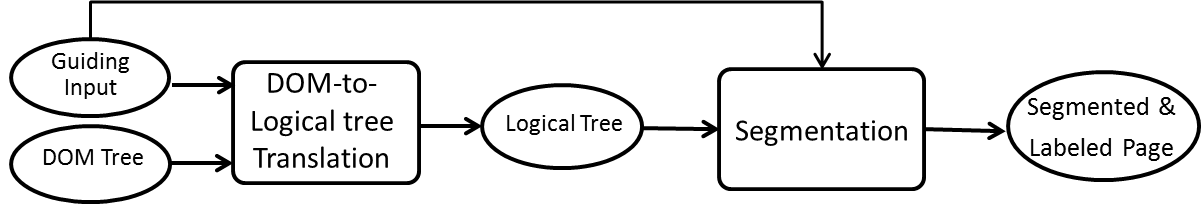}
\caption{Modified BoM segmentation model}
\label{ModifiedBoMmodel}
\end{figure}

\section{The Multiscreen-oriented-\\
segmentation Approach}
\label{sec:modifiedBoM}
\subsection{Overview}
\label{subSec:overModifiedBoM}
Our method segments a raw page based on input which guides: (1) the behavior analysis of DOM elements, (2) the labeling of logical leaf nodes with a function and (3) the production of blocks from logical leaf nodes. Specifically, the functions we use, i.e., "multimedia" and "interactive", are derived from the device features.
The following sections describe each phase of the MSoS approach that is depicted in Figure ~\ref{ModifiedBoMmodel}.

\subsection{DOM to partially-labeled logical tree}
\label{subSec:analysis}

The goal of this phase is to abstract the DOM tree and to represent the page in the form of a logical tree specific to our approach, in which we seek to label each node with a function and to minimize the number of leaf nodes to optimize the segmentation process. 

Similar to BOM, the DOM tree is first filtered and for each retained DOM element a logical node is created and added to the logical tree. 
To then label a logical node with a function, we analyze the static and dynamic behavior of the corresponding DOM element. 
The element behavior is identified through its tag name, its HTML attributes and its JS properties. In particular, we check the properties that can alter the static behavior of an HTML element, e.g., event listeners, and the HTML5 attributes that identify the role of an element, e.g., 'role'. If the function of a DOM element corresponds to one of the guiding input functions, we label the logical node with this function. There are elements whose behavior does not satisfy any of the functions. In this case, their corresponding logical node remains non-labeled. Thus, the leaf nodes of the resulting logical tree are not all labeled and their number is relatively big. 

To reduce their number, the logical tree is optimized to form geometrically bigger labeled blocks. 
This optimization will serve the next segmentation phase. 
The optimization procedure is as follows: (1) the tree is traversed from the root to the leaf nodes in a breadth-first manner. (2) If a node is labeled, we check whether its siblings are labeled with the same function. If positive, we merge them to form one node. (3) After analyzing all siblings, if only one labeled child remains, we propagate its label to its parent. 
The output of this first phase is a logical tree with a smaller number of nodes but with bigger geometry. It should be noted that some nodes may still be non-labelled.

\subsection{Segmentation}
\label{subSec:processing}
The segmentation consists of producing labeled blocks from the partially-labeled logical tree. A trivial segmentation that produces one block from each logical leaf node, results in creating an excessive number of blocks. A better segmentation can be obtained by (1) merging logical nodes according to the Gestalt laws and the pG value, (2) while keeping blocks with different functions separated and (3) making all leaf nodes labeled with the adequate function.

In this work, we consider the notions of global and local pG as defined in BoM\cite{bom}.   
The global pG is set before starting the segmentation.
The local pG is updated during runtime, as described in the next paragraph, to adapt the segmentation of the node subtree to its content. 
Both the global and local pG values are calculated automatically by considering the geometry of the labeled descendants respectively in the entire logical tree and in a local subtree as follows:
for all the labeled nodes, we calculate the ratio of their areas over the relevant page area. We define the relevant page area as the rectangular area defined by the top-left corner of the page, a width equal to the page width, and a height set to the minimum between the page height and five times the screen height.
We set the pG value of a subtree to the biggest pG value in this subtree, or to the global pG if the subtree does not have labeled descendants. Intuitively, the bigger the local pG is, the fewer final blocks will be produced and the better the segmentation results are. 

Then, we proceed with the processing of the logical tree in a depth-first manner starting from the root node and using the global pG value, as follows: 
(1) if a node is labeled, we try to merge it with its siblings, as described below.
(2) if a node is non-labeled and its descendants have different labels, we process its subtree.
(3) if a node is non-labeled and its descendants have only one function and its relative area is bigger than the pG, then we process its subtree; otherwise, we investigate the possibility of merging it with its siblings.

We try to merge a node with its next siblings, as follows: if the node does not have any sibling, it produces a block. Otherwise, for each sibling, if one of the functions of the sibling descendants is different from the current node function, the nodes are not merged even if the Gestalt laws and geometrical conditions are satisfied, and the current node produces a block. 
Otherwise, if the functions are the same, the merging of nodes is tested using the Gestalt laws and the geometrical conditions, as used in BoM. At the end of the merging, at least one labeled block is produced.

\section{Experimentation}
\label{sec:experimentation}

\begin{figure}[t!]
\centering
\includegraphics[height=1.3in,width=3.5in]{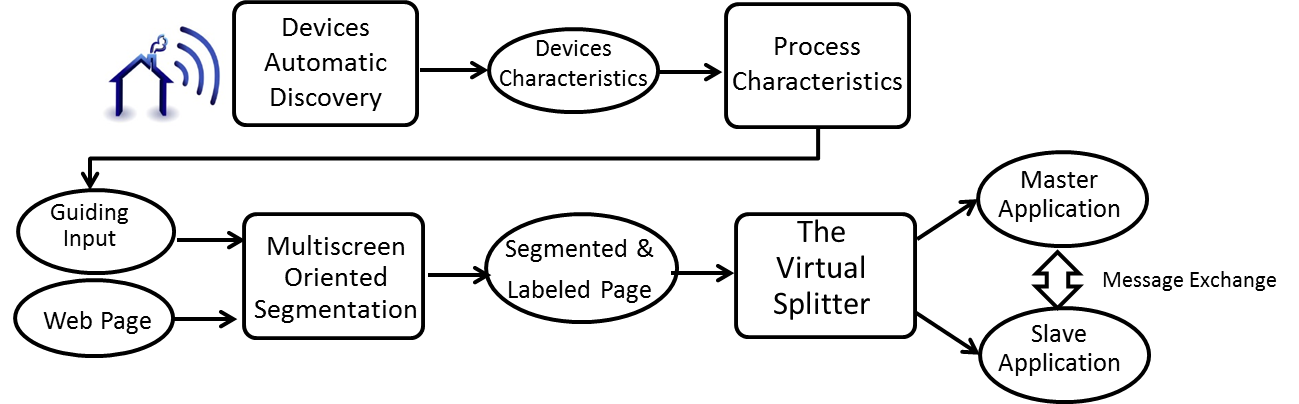}
\caption{The fully automatic virtual splitting system}
\label{enhancedVS}
\end{figure}
\subsection{Deployment and integration aspects}
\label{subSec:setup}

To deploy our approach and validate that it produces results that are useful in the context of multi-screen applications, we integrated it with the COLTRAM platform \cite{coltramEuroITV} and the VSplitter \cite{theVirtualSplitter}. 

We developed a COLTRAM web service to automatically discover devices available in the network and to get a list of their features. We limit the number of devices to two. We characterize each device based on: (1) its screen size, i.e., large or small display, (2) its means of interaction, i.e., touch input, keyboard, mouse or non-interactive, (3) its type, i.e., TV, portable device or desktop. Then, for each device we identify their dominant feature that we consider as the function of the device. For instance, a smartTV is better used for displaying "multimedia" content, e.g., image, video, etc., a smart phone with a touch-screen is adequate for "interaction" purposes. 
As depicted in Figure~\ref{enhancedVS} the guiding input, formed as a json object that associates each device with a function, and the application DOM tree are fed to MSoS.

Based on this input, we statically classify the element tags. For instance, the 'audio', 'video' and 'object' tags are used to embed multimedia content in a web document.
In the W3C content categories, the interactive content is limited to the HTML tags that are initially intended for the user interaction. In our classification, we did not adopt the interactive content definition given by W3C because:(1) some tags in the category are more multimedia than interactive, e.g., video and audio elements with a control bar, (2) some elements can become interactive after event listeners are registered on them to listen to user interaction events. 
In consequence, we analyze first the HTML attributes that are set statically in the HTML document, or dynamically on document load.
And second, we capture event listeners by instrumenting the addEventListener native method.

The VSplitter refactors single-screen applications, delivers a multi-screen application, and maintains the application functionality across devices by monitoring the application updates and synchronizing the content between two devices, the master and the slave. The VSplitter uses an annotated DOM tree with annotations indicating to which device a DOM element should belong. In order to annotate the application DOM tree, we exploit the fact that the logical nodes contain a reference to their corresponding DOM elements. We annotate DOM elements based on the label of their corresponding logical node. If this label refers to the selected function of the master device, then we annotate the element as 'device1'. Otherwise, if it refers to features of the slave device, we annotate it as 'device2'. Since the logical tree does not cover the complete DOM tree, but only the retained elements during the abstraction phase, the DOM tree is not totally annotated. The annotation is then resolved as denoted in our previous work \cite{theVirtualSplitter} and the content is distributed over the master and slave applications. Each of these applications is wrapped in a COLTRAM application and exposes a service for communication. Both master and slave applications are discoverable by each other, thus allowing a communication channel. Using this channel, updates and synchronization messages are exchanged continuously between the master and slave applications.
\subsection{Results and Discussion}
\label{subSec:results}
In this section, we illustrate the MSoS results by comparing them to BoM results and we evaluate our MSoS by comparing it to a ground truth that we refer to as GT. 
The procedure is based on the evaluation of three performance parameters: the visual coherence of blocks, the correctness of the function attributed to each block, and finally that blocks do not have content with different functions.

To test our method, we selected ten existing pages with multimedia and interactive content, classified as follows: (1) social applications i.e., YouTube, (2) video player applications, i.e., mediaElement\footnote{http://mediaelementjs.com/}, videojs\footnote{http://www.videojs.com/}, jplayer\footnote{http://jplayer.org/}, (3) web synchronized applications, e.g., semantic video\footnote{http://popcornjs.org/demo/semantic-video}. Applications and results are accessible from our site \footnote{http://download.tsi.telecom-paristech.fr/gpac/MSoS}.

\begin{figure}
\noindent\makebox{
\subfigure[]{
	\includegraphics[height=1.8in, width=0.22\textwidth]{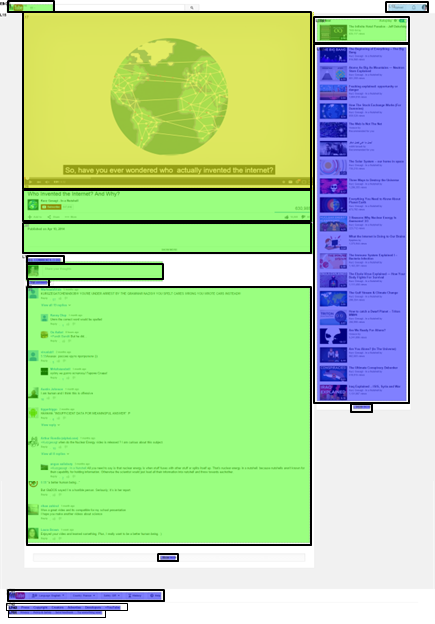}
 	\label{sf:BoM0.31}
}
\subfigure[]{
	\includegraphics[ height=1.8in, width=0.22\textwidth]{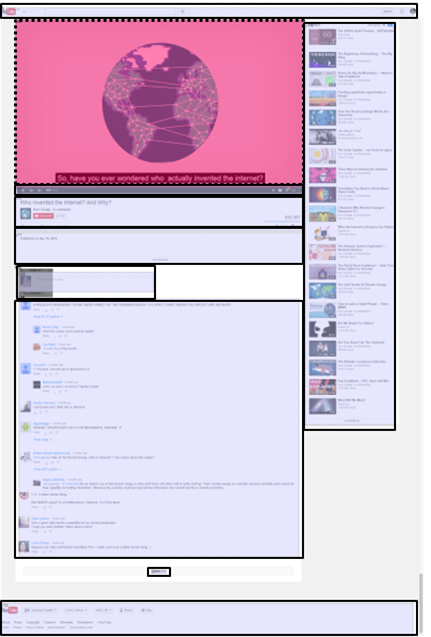}
 	\label{sf:adaptiveBoM}
}
\quad
}
\caption []
     { \label{fig:fullfig} Segmentation results on a YouTube page \subref{sf:BoM0.31}
      BoM with pG = 0.31     \subref{sf:adaptiveBoM} MSoS with pG = 0.31 and 0.36
    }
\label{fig:results}
\end{figure}

\textbf{Comparing to BoM:}
We illustrate the MSoS results by comparing them to the same page segmented by BoM.  Figure \ref{fig:results} presents the two segmentation results on a YouTube page \footnote{http://bit.ly/1eue6i3}. Note that we cropped the comments section to better illustrate the segmentation results. Figure \ref{sf:BoM0.31} represents the segmentation results using BoM with a pG value set manually to 0.31. 
Note here that the block colors are internal to BoM. Figure \ref{sf:adaptiveBoM} represents the segmentation results with MSoS. During the segmentation, two main values were computed: 0.36 (global and local) and 0.31 (local).
Most of the logical nodes were processed with the 0.31 value, this is why we decided to configure BoM with this value.
Comparing the two figures, the blocks generated from MSoS are more coherent than those of BoM, in particular, the header, footer and the sidebar section. Note that BoM did not consider the search bar on the top of the page, while our algorithm did.
In Figure \ref{sf:adaptiveBoM} the YouTube video controls are identified as a block separated from the video block. This proves that content functions were taken into consideration during the segmentation. The grey blocks refer to interactive blocks and the unique purple block refers to the multimedia content. Note that video subtitles are judged as multimedia content since they overlap the video element. Our method ensured the separation of blocks with different functionality, thus facilitating the content mapping to the 'best-match' device in the context of multi-screen environment.

\begin{table}[h]
	\tiny
	\begin{center}
       \begin{tabular}{| >{\centering\arraybackslash}m{0.4in} | >{\centering\arraybackslash}m{0.4in}
        	|>{\centering\arraybackslash}m{0.4in} | >{\centering\arraybackslash}m{0.4in} |
        	>{\centering\arraybackslash}m{0.4in} |}
\cline{1-5}
\multirow{2}{*}{Applications} & \multirow{2}{*}{Precision} & \multirow{2}{*}{Recall}& \multicolumn{2}{ c| }{Non-Matching} \\ \cline{4-5}
& & &  Over-Segmented & Non-Related\\ \cline{1-5}
\multicolumn{1}{ |c  }{\multirow{1}{*}{ YouTube pages } } &
\multicolumn{1}{ |c| }{0.38} &  0.67 & 0.08 & 0.46\\ [2pt]\cline{1-5}
\multicolumn{1}{ |c  }{\multirow{1}{*}{Video player pages} } &
\multicolumn{1}{ |c| }{0.74} & 0.80& 0.056 & 0.18\\[2pt] \cline{1-5}
\multicolumn{1}{ |c  }{Video Sync}                        &
\multicolumn{1}{ |c| }{0.72} & 0.96 & 0.035 & 0.21\\ [2pt]\cline{1-5}
	\end{tabular}
	\caption{Evaluation of the MSoS approach}
  \label{table:results}
  \end{center}
\end{table}

\textbf{Comparing to a ground truth:}
In order to evaluate our MSoS approach a GT was created manually, where coherent blocks were created and assigned a function between 'Multimedia' and 'Interactive'. Afterwards, we compared the MSoS results to the GT and we provide the comparison results in Table \ref{table:results} in the form of precision and recall rates. 
We define the precision and recall rates as follows: 
\\
$Precision = {Nb\ of\ Matching\ Blocks}/{Nb\ of\ MSoS\ Blocks}$,
\\
$Recall = {Nb\ of\ Matching\ Blocks}/{Nb\ of\ GT\ Blocks}$.\\
The Recall is equal to one if the MSoS could identify correctly all the blocks of the GT.
The Precision is equal to one if the MSoS did not produce any non-matching block.
The non-matching column refers to the average number of blocks that: 1) are over-segmented by the MSoS, 2) have no correspondence with any block in the GT or they are not correctly labeled.

As Table \ref{table:results} shows, the precision and recall rates for YouTube pages are the lowest (0.38 and 0.67 resp.). This is due to the high average of non-related blocks. These non-related blocks are due to the identification of additional interactive blocks that are present in YouTube pages but are not visible for users, e.g., the guiding block that appears when we click on the button next to the YouTube logo in the top of the page. 
The tests conducted on the applications from video-player libraries always lead to the separation of the control bar from the video element. This is validated by the high precision and recall values (0.74 and 0.80 respectively). The non-related blocks here refer to the cases where the MSoS could not label some blocks since they were not merged with any labeled blocks. In addition in one application the subtitles were merged with the control bar while they should have been merged with the video element.
For the third set of applications, the precision rate is 0.96 and it indicates that almost all the blocks of the GT were identified by MSoS.
Though the average number of the over-segmented blocks is small for the three set of applications (0.08, 0.056 and 0.035 resp.), it is important to note here that the over-segmentation is the drawbacks of using a pG value that is calculated according only to the labeled nodes.
\section{Conclusion}
This paper proposed a multiscreen-oriented segmentation approach. The segmentation method is hybrid and aims at creating coherent blocks and separating blocks of different functionalities given as a guiding input. MSoS is completely automatic and characterized by a dynamic adaptation to the page content. It is inspired by an abstraction model proposed in BoM. To validate our work, MSoS was integrated within the virtual splitting system for application distribution in the multi-screen context. Experiments were conducted on a set of existing multimedia applications.
We compared the MSoS results to BoM\cite{bom} and to a manually created ground truth. With our adaptive method, better segmentation results were obtained especially in critical regions of the page and blocks with different function were kept separated. 
With a 81\%\ precision, the MSoS is a promising method. 
As a perspective, we are planning to enlarge our application dataset to compare our MSoS qualitatively and quantitatively to different segmentation methods. In addition, we want to extend the evaluation plan to consider the importance of the block position on block identification.
\label{sec:conclusion}

\bibliographystyle{plain}
\bibliography{bib/SemanticBOM}

\begin{thebibliography}{1}

\bibitem{vips}
D.~Cai, S.~Yu, J.R. Wen, and W.Y. Ma.
\newblock Vips: A vision-based page segmentation algorithm.
\newblock Technical report, Microsoft, MSR-TR-2003-79, 2003.

\bibitem{FOM}
J.~Chen, B.~Zhou, J.~Shi, H.~Zhang, and Q.~Fengwu.
\newblock Function-based object model towards website adaptation.
\newblock In {\em Proceedings of the 10th International Conference on World
  Wide Web}, WWW '01, pages 587--596, New York, NY, USA, 2001. ACM.

\bibitem{coltramEuroITV}
J.C. Dufourd, M.~Tritschler, L.~Bassbouss, R.~Bouazizi, and S.~Steglich.
\newblock An open platform for multiscreen services.
\newblock In {\em the 11th European Interactive TV conference EuroITV}, Como,
  Italy, June 2013.

\bibitem{bom}
A.~Sanoja and S.~Gan{\c{c}}arski.
\newblock Block-o-matic: A web page segmentation framework.
\newblock In {\em Multimedia Computing and Systems (ICMCS), 2014 International
  Conference on}, pages 595--600. IEEE, 2014.

\bibitem{theVirtualSplitter}
M.~Sarkis, C.~Concolato, and J.C. Dufourd.
\newblock The virtual splitter: Refactoring web applications for the
  multiscreen environment.
\newblock In {\em Proceedings of the 2014 ACM Symposium on Document
  Engineering}, DocEng '14, pages 139--142, New York, NY, USA, 2014. ACM.

\end{thebibliography}
\end{document}